\documentclass[twocolumn,preprintnumbers,prb]{revtex4}
\usepackage{graphicx}%
\usepackage{dcolumn}
\usepackage{amsmath}
\usepackage{amssymb}
\usepackage{bm}
\makeatletter
\def\btt#1{\texttt{\@backslashchar#1}}%
\DeclareRobustCommand\bblash{\btt{\@backslashchar}}%
\makeatother

\begin{document}

\preprint{MgAlB2.tex}

\title{Crystal Structure of the Mg$_{1-x}$Al$_x$B$_2$ superconductors near $x \approx$ 0.5}
\author{Serena Margadonna,$^1$ Kosmas Prassides,$^{2,3}$ Ioannis Arvanitidis,$^2$ Michael Pissas,$^3$ \\
Georgios Papavassiliou,$^3$ and Andrew N. Fitch$^4$}

\affiliation{$^1$Department of Chemistry, University of Cambridge, Cambridge CB2 1EW, UK\\
$^2$School of Chemistry, Physics and Environmental Science, University of Sussex, Brighton BN1 9QJ, UK\\
$^3$Institute of Materials Science, NCSR Demokritos, 153 10 Aghia Paraskevi, Athens, Greece\\
$^4$European Synchrotron Radiation Facility, BP 220, 38042 Grenoble, France}

\date{28 March 2002}
\begin{abstract}
Precise structural information on the Mg$_{1-x}$Al$_x$B$_2$ superconductors in the vicinity of 
$x \approx$ 0.5 is derived from high-resolution synchrotron 
X-ray powder diffraction measurements. We find that a hexagonal superstructure, accompanied by doubling 
of the $c$-axis, ordering of Mg 
and Al in alternating layers, and a shift of the B layers towards Al by $\sim$0.12 \AA, is formed. The 
unusually large width of the (00$\frac{1}{2}$) superlattice 
peak implies the presence of microstrain broadening, arising from anisotropic stacking 
of Al and Mg layers and/or structural 
modulations within the $ab$ plane. The ordered phase survives only over a limited range 
of compositions away from the optimum $x$= 0.5 doping level.
\end{abstract}
\maketitle

The discovery of superconductivity in the binary boride, MgB$_2$ at the high temperature of 39 K \cite{nagamatsu01} has 
generated considerable interest because of the apparent simplicity of its chemical composition, crystal 
structure and electronic properties. MgB$_2$ possesses a simple hexagonal structure (AlB$_2$-type, space 
group $P$6/$mmm$) comprising graphitic-type B layers interleaved with Mg layers \cite{jones}. Band structure 
calculations reveal that, while strong covalent B-B bonds are retained, 
Mg is fully ionized \cite{An01}. The charge carriers 
are situated in two essentially two-dimensional (2D) bands derived from the $\sigma$-bonding $p_{x,y}$-orbitals of boron, 
and in one electron and one hole band derived from the $\pi$-bonding $p_z$-orbitals of boron. There is considerable 
experimental evidence (boron isotope effect \cite{budko01a}, scanning tunnelling 
experiments \cite{STM}, negative pressure coefficient of $T_c$ \cite{pressure}) that 
a conventional phonon-mediated pairing mechanism can account for the superconducting 
properties of MgB$_2$, in which a key role is played by the 2D $\sigma$-band of 
$p_{x,y}$-orbitals within the boron layers. 
Consistent with this, a small discontinuity only in the boron interlayer spacing was observed 
at $T_c$ by precise structural measurements \cite{serena01}.

Changes in carrier concentration and their influence on the electronic properties 
of the system can provide crucial tests for the mechanism of superconductivity 
and they have been investigated through chemical substitution between \cite{slusky,zao} 
or within \cite{takenobu} the boron layers. In all cases, the critical 
temperature of MgB$_2$ decreases at various rates for different substitutions. 
Among the various ternary compositions, the properties of the Al-doped 
series Mg$_{1-x}$Al$_x$B$_2$ are of particular interest and have been extensively 
studied both experimentally and theoretically. Electron doping through 
Al substitution leads to a decrease in $T_c$ which can be rationalized by the 
filling of the electronic states by the additional electron donated by Al 
and the resulting decrease in the density of states at the Fermi level \cite{An01}. 
However, very importantly the rate of decrease of $T_c$, d$T_c$/d$x$ sensitively 
depends on the doping level, $x$. \cite{slusky} $T_c$ first decreases smoothly in the region 
0$< x <$ 0.1, then the transition becomes broader up to $x$= 0.25. 
For compositions with 0.25$< x <$ 0.4, $T_c$ drops more sharply and then 
superconductivity vanishes in the vicinity of $x$= 0.6 \cite{postorino,renker,li}. These 
results imply a more complicated crystal structural and/or electronic 
response for the Mg$_{1-x}$Al$_x$B$_2$ series than that expected for solid solution 
and rigid band behavior. Indeed early X-ray diffraction measurements have 
revealed the presence of structural anomalies associated with a miscibility 
gap and multiphase behavior in the regions 0.1$< x <$ 0.25 and 0.7$< x <$ 0.8.\cite{slusky}  
In addition, electron diffraction and transmission electron microscopy 
studies have provided direct evidence for the existence of a 
superstructure at $x \approx$ 0.5 resulting from ordered arrangements of Al and Mg atoms 
both along the $c$-axis \cite{li,zand} and in the $a-b$ plane \cite{zand}. 

In this paper, we report a structural determination of the 
Mg$_{1-x}$Al$_x$B$_2$ ($x$= 0.45, 0.5, 0.55) ternary superconductors by synchrotron X-ray 
powder diffraction at 16 and 298 K. A hexagonal superstructure, accompanied by the doubling 
of the $c$-axis of the MgB$_2$ structure is observed, arising from ordering of 
Al and Mg in subsequent layers. Formation of the superstructure is optimal for 
Mg$_{0.5}$Al$_{0.5}$B$_2$ but survives small deviations ($\approx$10\%) from the $x$= 0.5 composition. 

Powder samples with nominal composition Mg$_{1-x}$Al$_x$B$_2$ with $x$= 0.4, 0.45, 0.5, 
and 0.55 were prepared by liquid-vapor to solid reactions for 24 hours, as described in ref. \cite{pissas}. 
The reaction temperatures were between 800$^{\circ}$ and 870$^{\circ}$C. 
SQUID measurements were performed on 50-mg samples in the temperature range 1.8-50 K with a 
Quantum Design SQUID magnetometer (MPMS5). High-resolution synchrotron X-ray diffraction 
experiments were carried out on the BM16 beamline at the European Synchrotron Radiation Facility 
(ESRF), France. The samples were sealed in 1.0-mm diameter glass capillaries and diffraction profiles 
($\lambda$= 0.85023 \AA) were collected at 16 K (and for $x$= 0.5 also at ambient temperature) 
in continuous scanning mode using nine Ge(111) analyzer crystals. The capillaries were 
continuously spun during data acquisition. The data were rebinned in the 2$\theta$ range 
1-80$^{\circ}$ to a step of 0.01$^{\circ}$ and refined using the GSAS suite of Rietveld analysis 
programs. The peakshape of the diffraction lines was modelled by a convolution of a 
pseudo-Voigt function and an asymmetry function, which is related to the instrumental axial 
divergence \cite{cox}. In addition, in order to account for anisotropic peak broadening 
of different classes of ($hkl$) reflections evident in the high-resolution diffraction profiles, 
the Gaussian and Lorentzian portions of the peakshape function ascribed to microstrain broadening 
($\sigma ^2 _s$ and $\gamma_s$) were described by the semi-empirical Stephens 
formalism \cite{stephens}. In this model, the width of each reflection can be expressed in terms 
of moments of a multi-dimensional distribution of lattice metric parameters and can be related to 
distributions of elastic strains caused by defects or dislocations. 

Fig. 1 shows the results of the magnetic measurements at 10 Oe (ZFC conditions) for the 
Mg$_{1-x}$Al$_x$B$_2$ compositions with $x$= 0.45, 0.5, and 0.55. Diamagnetic shielding is evident 
at low temperature in all three samples. The one-step transitions are sharp and the transition 
temperatures $T_c$, defined by the intersection of line extrapolations below and 
above $T_c$, are 6, 3, and 4 K for $x$= 0.45, 0.5, and 0.55, respectively. Although the shielding 
fractions at 2 K are small, the sharpness of the transitions implies that superconductivity 
is of bulk nature and does not arise from sample inhomogeneities and composition fluctuations.

Inspection of the synchrotron X-ray diffraction profile of Mg$_{0.5}$Al$_{0.5}$B$_2$ at 16 K 
shows that all reflections can be assigned either to a hexagonal cell (space group $P$6/$mmm$ - AlB$_2$ type) 
with lattice constants $a \approx$ 3.045 \AA\ and $c \approx$ 3.506 \AA\ or to a small fraction ($<$1\%) of MgO impurity. 
However, an additional weak peak not attributable to an impurity phase is clearly visible at 
2$\theta$= 7.26$^{\circ}$ (Fig. 2). This extra reflection whose intensity is $\sim$26\% of the weak (001) 
reflection indexes as (00$\frac{1}{2}$) on the hexagonal unit cell and provides the unambiguous 
signature of the formation of a superstructure whose unit cell is exactly doubled along the $c$-axis. 
No other prominent peaks that could arise from the formation of the superstructure are observed under 
the present experimental conditions. The observed doubling of the $c$-axis allows ordering of Al and 
Mg in subsequent layers while maintaining the gross features of the original AlB$_2$-type structure. 
The simplest Al/Mg ordering motif consistent with the observed superreflection corresponds 
to an alternation of Al and Mg along the $c$-axis (space group $P$6/$mmm$), as suggested by TEM 
investigations in ref. \cite{li}. Rietveld refinements using this model proceeded smoothly with Mg 
and Al placed in the 1$a$ (0,0,$\frac{1}{2}$) and 1$b$ (0,0,0) positions, respectively, and the 
B atoms located in the 4$h$ ($\frac{1}{3}$,$\frac{2}{3}$,$z$; $z\approx$$\frac{1}{4}$) positions. 
The refined lattice parameters of Mg$_{0.5}$Al$_{0.5}$B$_2$ at 16 K are $a$= 3.04436(2) \AA\ 
and $c$= 6.71248(10) \AA. In addition 
to Al/Mg ordering, the present structural model allows for unequal separation between 
neighboring MgB$_2$ and AlB$_2$ slabs as the B atoms have the freedom to relax along the $c$-direction 
($z \neq$$\frac{1}{4}$), while the strictly flat boron sheets of the original AlB$_2$-type structure 
are maintained (Fig. 3). In Mg$_{0.5}$Al$_{0.5}$B$_2$, the B position refines to 
($\frac{1}{3}$,$\frac{2}{3}$,0.2413(3)), corresponding to Mg-B and Al-B distances of 
2.471(1) \AA\ and 2.390(1) \AA, respectively. 

In the course of the Rietveld refinements, it was evident that the widths of the (00$l$) reflections 
were invariably larger than those of the ($hk$0) reflections, providing the signature of strong anisotropic 
strain broadening effects. These microstructural effects were modeled with the formalism developed by
Stephens \cite{stephens}. For this model, the obtained microstrain broadening contribution to the width of the 
(001) reflection is $\sim$1.3\%, almost twice as large as that to the (200) peak. This could be 
related to increased strains along the $c$-axis due to the presence of defects and stacking faults 
associated with the Al and Mg layers. Moreover, the microstrain contribution to the width of the (00$\frac{1}{2}$) 
superlattice peak is even higher ($\sim$4.1\%) implying an anisotropic distribution along the $c$-axis of 
alternating ordered Al and Mg layers. Fig. 4 displays the final refinement of the synchrotron X-ray 
diffraction profile of Mg$_{0.5}$Al$_{0.5}$B$_2$ at 16 K, while the results of the analysis are summarized in Table I. 

Despite the excellent quality of the Rietveld refinement of the synchrotron X-ray diffraction profile 
of Mg$_{0.5}$Al$_{0.5}$B$_2$ using the Al/Mg ordering model described above, alternative models 
of ordering were also considered. An attractive possibility consistent with the observed doubling 
of the hexagonal $c$-axis involves ordering of both Al and Mg on the same layer with each atom 
surrounded by as many atoms of the other kind as possible. However, the resulting crystal 
symmetry is $P$6$_3$/$mmc$ and the (00$\frac{1}{2}$) reflection will be extinct, clearly necessitating 
ordering of Al and Mg atoms in alternating layers. Finally, we considered the possibility that 
ordering of Mg and Al is accompanied by distortion of the B layers. The appropriate unit cell 
in this case is primitive orthorhombic with lattice constants $a \approx$ $a_h$, 
$b \approx$ $b_h$$\sqrt{3}$, and $c$= 2$c_h$. 
Rietveld refinements of the diffraction data using such models did proceed to convergence and improved 
quality of fit but the paucity of observed superreflections did not allow unambiguous conclusions 
concerning the existence of an orthorhombic distortion of the original hexagonal unit cell. 

A synchrotron X-ray diffraction profile of Mg$_{0.5}$Al$_{0.5}$B$_2$ was also collected at 298 K. The 
(00$\frac{1}{2}$) superlattice peak is again present implying the absence of a phase change with 
increasing temperature. Rietveld refinement with the $P$6/$mmm$ superstructure model proceeded smoothly 
(Table I), leading to lattice constants of $a$= 3.04705(2) \AA\ 
and $c$= 6.72409(12) \AA\ and anisotropic expansion between 16 and 298 K of 0.088(1)\% and 0.173(3)\% 
along the $a$- and $c$-axis, respectively.

Synchrotron X-ray diffraction profiles were also collected at 16 K for the 
Mg$_{0.55}$Al$_{0.45}$B$_2$ and Mg$_{0.45}$Al$_{0.55}$B$_2$ compositions. In both cases, 
the peak at 2$\theta \approx$ 7.26$^{\circ}$ which provides the signature of Mg/Al ordering and superstructure 
formation is present but with decreased intensity ($\approx$11\% and 18\% of the corresponding (001) reflections, 
respectively) and increased width when compared to that in Mg$_{0.5}$Al$_{0.5}$B$_2$ (Fig. 2). The 
two diffraction profiles were refined using the same structural model discussed above. In the 
case of Mg$_{0.45}$Al$_{0.55}$B$_2$, the excess Mg (5\%) present was disordered in the Al layers, 
while for Mg$_{0.45}$Al$_{0.55}$B$_2$, the excess Al (5\%) in the Mg layers. The Rietveld 
refinements proceeded smoothly and the refined values of the lattice constants are 
$a$= 3.04318(3) \AA, $c$= 6.69127(12) \AA\ and $a$= 3.04988(2) \AA, $c$= 6.73190(11) \AA\ for 
Mg$_{0.45}$Al$_{0.55}$B$_2$ and Mg$_{0.55}$Al$_{0.45}$B$_2$, respectively (Table II). 

A perspective view of the hexagonal superstructure of Mg$_{0.5}$Al$_{0.5}$B$_2$ is shown in Fig. 3. 
Compared to the unit cell of MgB$_2$, the present structure arises from ordering of the 
Mg$^{2+}$ and Al$^{3+}$ ions, leading to doubling of the unit cell along the $c$ axis. Both 
crystallographically distinct Mg$^{2+}$ and Al$^{3+}$ ions present in the unit cell have identical 
coordination environments, namely they lie directly above the centers of two B hexagons of adjacent 
B layers. Consistent with the higher ionic charge and smaller ionic radius (0.675 \AA) of Al$^{3+}$ 
compared to Mg$^{2+}$ (0.860 \AA), the Al-B bond distances are smaller than the Mg-B ones, 
2.390(1) \AA\ and 2.471(1) \AA, respectively at 16 K. This leads to Al-B and Mg-B layer 
separations along the $c$-direction of 1.620(2) \AA\ and 1.737(2) \AA, respectively, and reflects a 
displacement of the B layers of $\sim$0.12 \AA\ towards each Al layer. As we deviate from the 
$x$= 0.5 ternary towards Mg or Al rich compositions, the difference between the Al-B and Mg-B 
layer separations rapidly decreases (only $\sim$0.01 and 0.05 \AA, respectively, for $x$= 0.45 and 0.55, 
Table II). The ordered superstructure is essentially destroyed for Mg$_{0.6}$Al$_{0.4}$B$_2$, 
as the (00$\frac{1}{2}$) superlattice peak has now collapsed into the background (Fig. 2), 
while further it is unambiguously absent in compositions with even smaller Al doping levels. 
On the other hand, the in-plane B-B bond lengths decrease monotonically with increasing Al 
content across the stability boundary of the superstructure (from 1.76365(1) \AA\ in 
Mg$_{0.6}$Al$_{0.4}$B$_2$ to 1.75692(1) \AA\ in Mg$_{0.45}$Al$_{0.55}$B$_2$) in agreement with 
the strengthening of the in-plane $\sigma$-bonds.

The superstructure derived in this work is identical to that proposed by earlier TEM 
measurements \cite{li} and is energetically preferred, according to recent theoretical 
calculations \cite{calc}. In addition, a more complicated Mg/Al ordering scheme, involving 
both ordering along the $c$-axis and a sinusoidal modulation component, $q$ in the hexagonal $ab$ plane 
was observed in the HREM work of Zandbergen $et$ $al$ \cite{zand}. The signature of the in-plane structural 
modulation came from the splitting of the observed superreflections which yielded diffraction rings. 
In the present powder X-ray diffraction experiments, no such clear splittings are observed. 
However, we first note that the observed (00$\frac{1}{2}$) superreflection is unusually broad (Fig. 2) 
and the large microstrain broadening ($vide$ $supra$) required to account for its width may be precisely 
associated with such ordered distributions of the Mg and Al atoms within the hexagonal planes. 
In addition, there is a shoulder on the high angle side of the (00$\frac{1}{2}$) peak at 2$\theta \approx$ 
7.5$^{\circ}$ (marked * in Fig. 2) which indexes as ($q_1$,$q_2$,$\frac{1}{2}$) with 
($q_1^2$+$q_2^2$)$^{\frac{1}{2}}$$\approx$ 0.12, tantalizingly close to the incommensurate in-plane 
modulation vector, $q \approx$ 0.1 in ref \cite{zand}. 

The origin of the observed superstructure is of particular importance and may have consequences 
for the understanding of superconductivity in these systems. One possibility is that it is associated 
with an electronic instability near the 50\% doping level \cite{zand}. This is consistent with the 
extremely narrow stability range of the superstructure and its sensitivity to small compositional changes. 
On the other hand, the high quality of the present Mg$_{0.5}$Al$_{0.5}$B$_2$ sample precludes the 
presence of large compositional fluctuations that could be responsible for the observation of 
diamagnetic shielding and provides evidence that the superstructure does not suppress superconductivity, 
as it might be expected from the opening of a gap at the Fermi surface. Another possibility 
is that it is associated with a structural instability arising from the size mismatch between Al and 
Mg. The Al-B and Mg-B bond distances are 2.375 and 2.501 \AA\ in the AlB$_2$ and MgB$_2$ end members, 
respectively.  Doping of AlB$_2$ with Mg or MgB$_2$ with Al has only a small effect on the in-plane 
lattice constants but affects principally the interlayer separations. In both cases, increased doping 
is accompanied by the appearance of phase separation at certain critical 
levels \cite{slusky,postorino,li}. The coexisting phases differ mainly in their $c$-lattice constants 
and contain different concentrations of Mg and Al in the metal layers. In the vicinity of $x \approx$ 0.5, 
the size mismatch effect is maximal and complete ordering of Mg and Al becomes energetically favourable. 
The observed Al-B and Mg-B bond distances of 2.390(1) \AA\ and 2.471(1) \AA\ in the 
Mg$_{0.5}$Al$_{0.5}$B$_2$ superstructure are comparable to those encountered for the $x \approx$ 0.8 and 0.2 
compositions, respectively, near the Al- and Mg-rich critical concentrations for the two-phase separation 
onsets in the Mg$_{1-x}$Al$_x$B$_2$ series.

In conclusion, a hexagonal superstructure is obtained in Mg$_{1-x}$Al$_x$B$_2$ for a small range of 
Al and Mg concentrations near $x \approx$ 0.5. The principal component of the structural modulation of 
the parent MgB$_2$ structure is along the $c$-axis, while evidence exists for anisotropic distributions 
of alternating Mg/Al planes in the $c$ direction or ordered Mg/Al distributions within the 
hexagonal planes. Size mismatch and electronic effects were considered as possible 
origins of the observed behavior.

We thank the ESRF for provision of synchrotron X-ray beamtime. SM thanks Jesus College, Cambridge 
for a Research Fellowship. The research has been supported by a Marie Curie Fellowship of the 
EU program "Improving the Human Research Potential" under contract number HPMF-CT-2001-01436.

\begin{table}[h]
\small
\caption{Refined parameters for Mg$_{0.5}$Al$_{0.5}$B$_2$ obtained
from Rietveld refinement of synchrotron X-ray powder diffraction data at 16 K (space group 
$P$6/$mmm$; hexagonal cell constants: $a$= 3.04436(2) \AA, $c$= 6.71248(10) \AA; reliability factors: 
$R_{wp}$= 5.4\%, $R_{exp}$= 4.0\%). The lattice constants at 298 K are $a$= 3.04705(2) \AA, $c$= 
6.72409(12) \AA\ and the reliability factors $R_{wp}$= 6.1\%, $R_{exp}$= 5.1\%.}
\vspace{3mm}
\begin{tabular}{|c|c|c|c|c|c|c|}\hline
Atom & Site  & $x$/$a$ &  $y$/$a$ &  $z$/$c$  & $B$ (\AA$^2$) & occupancy \\
\hline\hline
Mg & 1$a$ & 0 & 0 & $\frac{1}{2}$ & 0.17(1) & 1.0(1) \\
Al & 1$b$ & 0 & 0 & 0 & 0.17(1) & 1.0(1) \\
B & 4$h$ & $\frac{1}{3}$ & $\frac{2}{3}$ & 0.2413(3) & 0.93(3) & 1.1(2) \\
\hline
\multicolumn{7}{l}{ \footnotesize}
\end{tabular}
\end{table}

\begin{table}[h]
\small
\caption{Selected bond distances (\AA) in Mg$_{1-x}$Al$_x$B$_2$ ($x$= 0.4, 0.45, 0.5, 0.55) at 16 K.}
\vspace{3mm}
\begin{tabular}{|c|c|c|c|c|}\hline
$x$ & Mg-B  & Al-B &  (Mg/Al)-B &  B-B  \\
\hline\hline
0.4 & - & - & 2.44157(2) & 1.76365(1)  \\
0.45 & 2.438(2) & 2.433(2) & - & 1.76085(1) \\
0.5 & 2.471(1) & 2.390(1) & - & 1.75766(1) \\
0.55 & 2.442(2) & 2.410(2) & - & 1.75692(1) \\
\hline
\multicolumn{5}{l}{ \footnotesize}
\end{tabular}
\end{table}

\vspace{3mm}
{\bf Figure captions}

\vspace{3mm}
{\bf Figure 1.} Temperature dependence of the magnetic susceptibilities of Mg$_{1-x}$Al$_x$B$_2$ 
($x$= 0.45, 0.5, and 0.55) measured under ZFC conditions in a field of 10 Oe. 

\vspace{3mm}
{\bf Figure 2.} Synchrotron X-ray powder diffraction profiles of 
Mg$_{1-x}$Al$_x$B$_2$ ($x$= 0.4, 0.45, 0.5, and 0.55) at 16 K in the vicinity of the 
(00$\frac{1}{2}$) superlattice reflection. The solid lines are guides to the eye.

\vspace{3mm}
{\bf Figure 3.} Structural model of the Mg$_{0.5}$Al$_{0.5}$B$_2$ superstructure.

\vspace{3mm}
{\bf Figure 4.} Final observed (points) and calculated (solid line) synchrotron X-ray powder diffraction 
profiles for Mg$_{0.5}$Al$_{0.5}$B$_2$ at 16 K in the range 6$^{\circ}$ to 80$^{\circ}$ 
($\lambda$= 0.85023 \AA). The lower panel shows the difference profiles and the ticks 
mark the positions of the Bragg reflections of Mg$_{0.5}$Al$_{0.5}$B$_2$ (upper) and MgO 
($<$1\%, lower). Inset. Rietveld fit in the range 6$^{\circ}$ to 15.6$^{\circ}$.

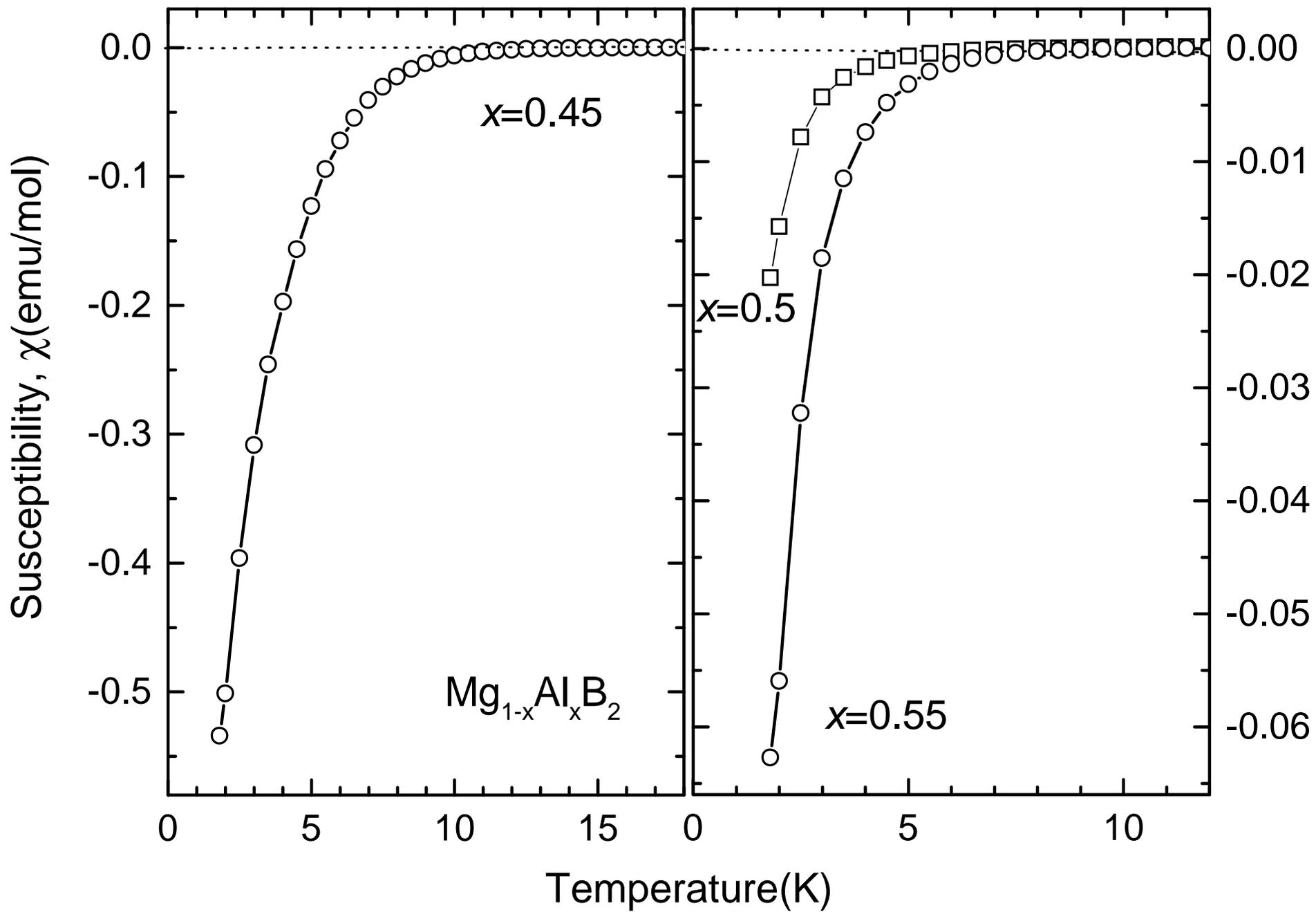
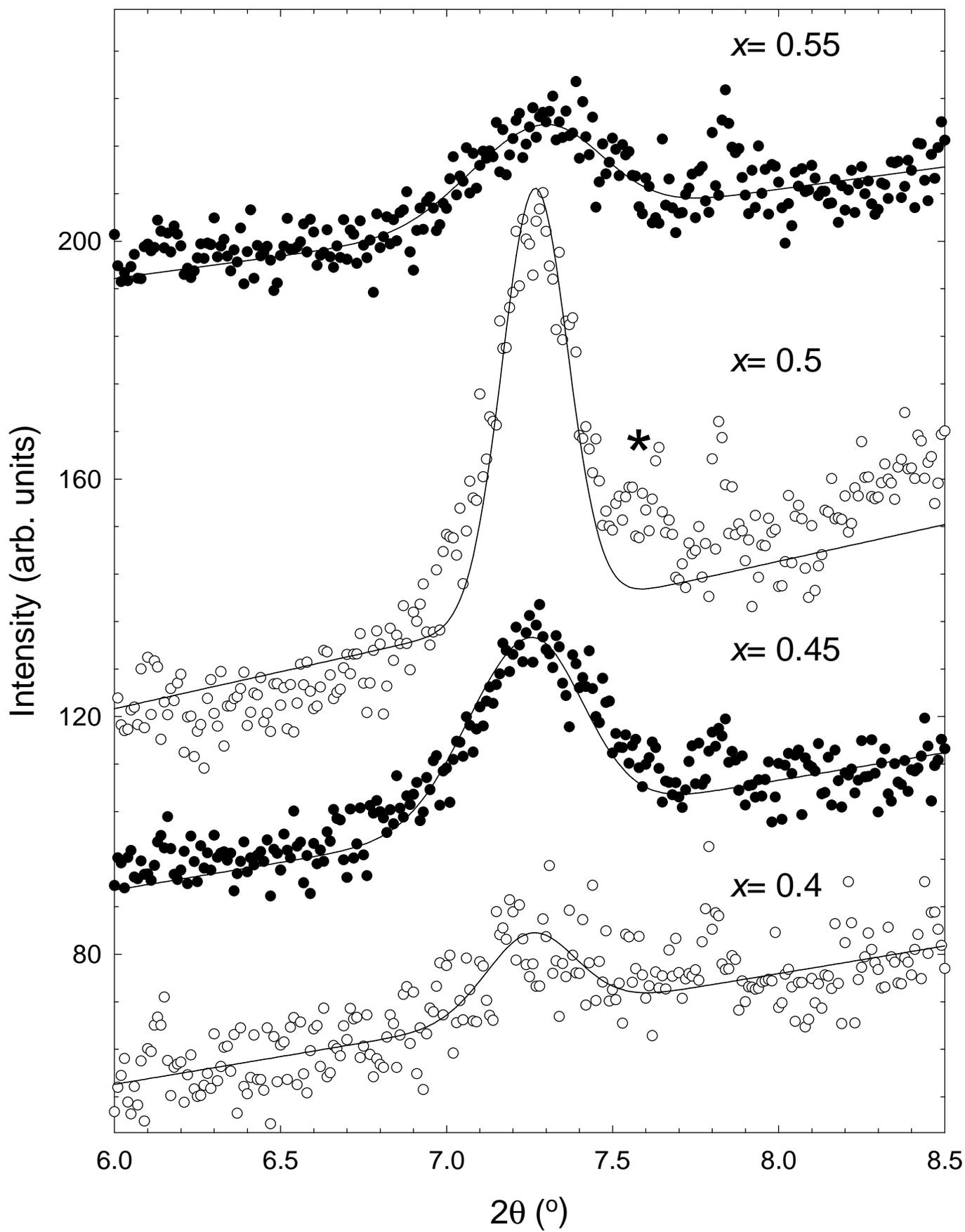
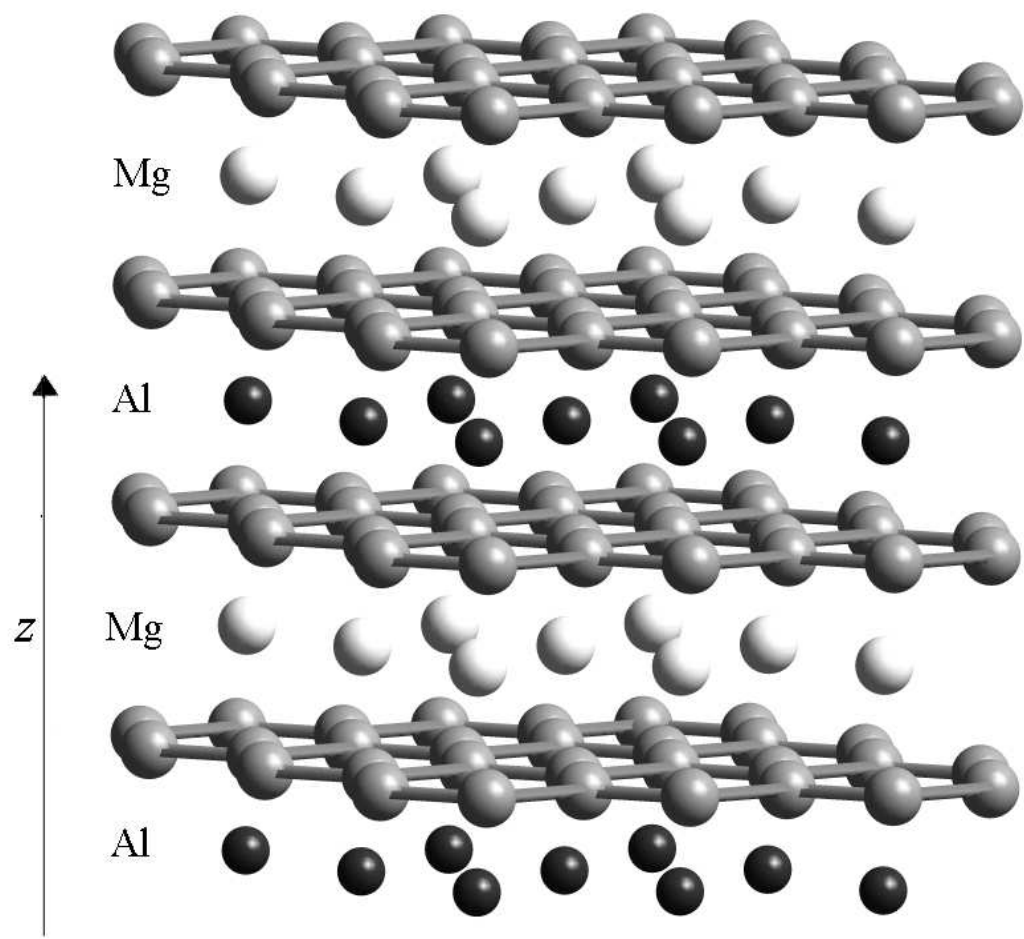
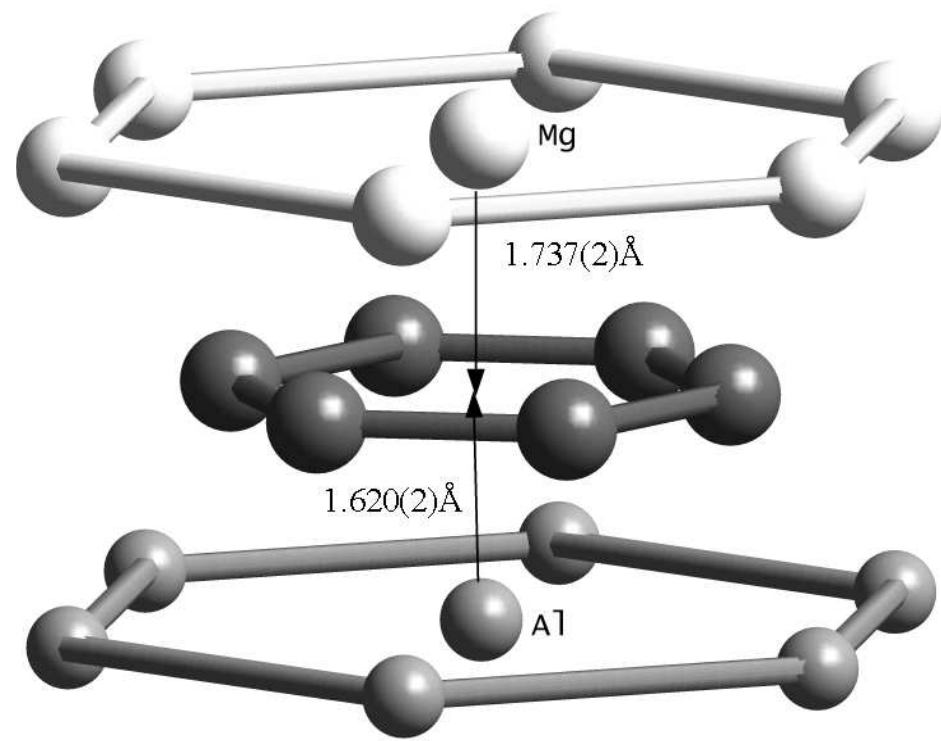
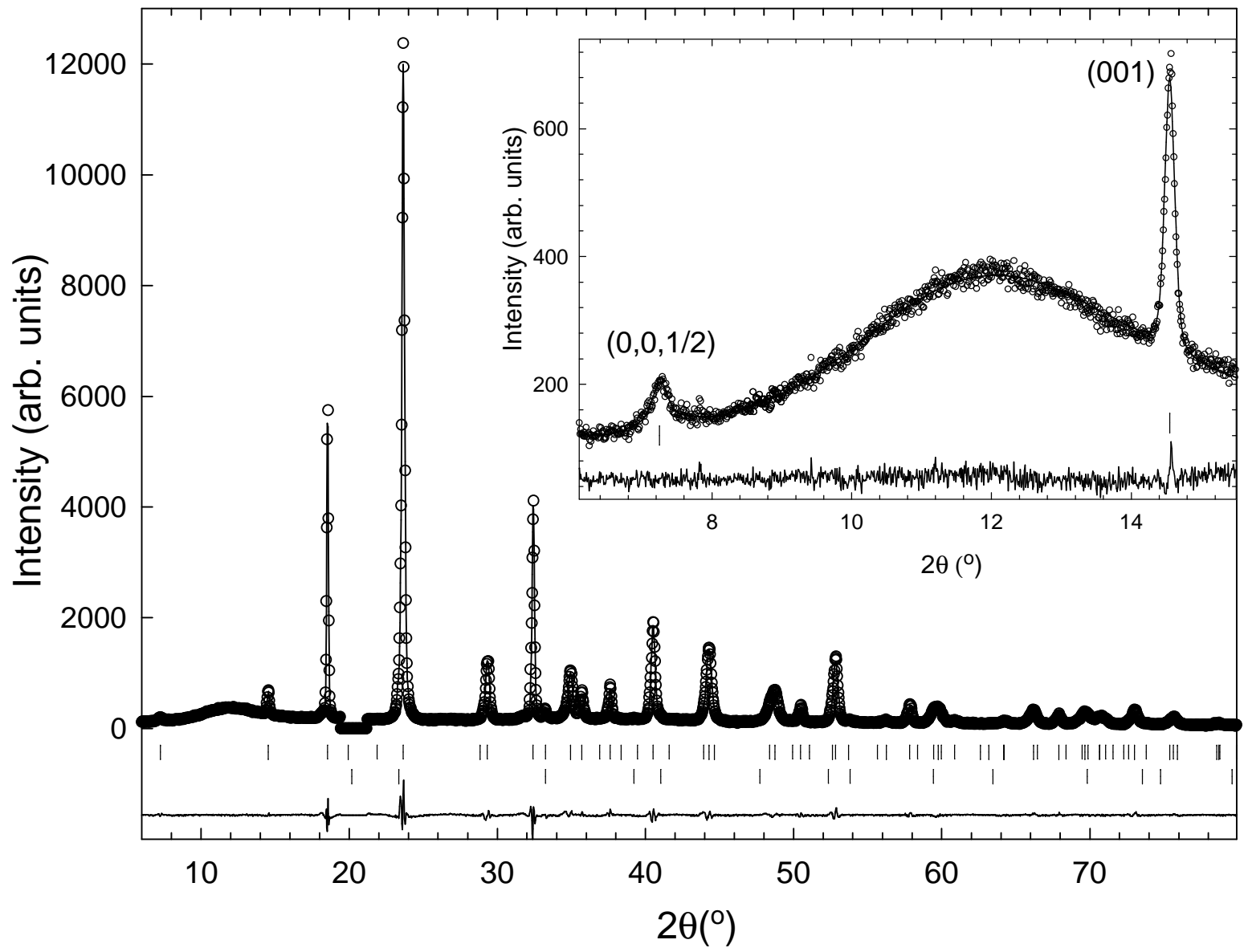
\end{document}